\documentclass[twocolumn]{revtex4}
\usepackage{amsmath}
\usepackage{amssymb}
\usepackage{rsfs}
\usepackage{color}
\usepackage{ulem}
\newcommand{\ra}{\rangle}

\begin{document}

\title{De Sitter Spacetime from Holographic Flat Spacetime with Inexact Bulk Quantum Mechanics}

\author{Eiji Konishi}
\email{konishi.eiji.27c@kyoto-u.jp}
\address{Graduate School of Human and Environmental Studies, Kyoto University, Kyoto 606-8501, Japan}

\date{\today}

\begin{abstract}
We argue that the flat spacetime with inexact quantum mechanics in it is dual to the de Sitter spacetime with exact quantum mechanics in it, and the positive cosmological constant of this de Sitter spacetime is in the second order of the degree of the violation of the bulk quantum mechanics in the flat spacetime.
The flat spacetime is holographic and has a dual time-contracted boundary conformal field theory with two redefined central charges at null infinity.
The vanishing smallness of the observed positive cosmological constant suggests the extraordinary exactness of the bulk quantum mechanics in the flat spacetime.
\end{abstract}

\maketitle

\section{Introduction}

The holographic principle asserts that the degrees of freedom (DoF) of a bulk space are encoded in the boundary quantum field system as information \cite{Hol1,Hol2,Hol3}.
The known examples of this principle are the black hole entropy \cite{B,H,GH,SV} and the $d+2$-dimensional anti-de Sitter spacetime/$d+1$-dimensional conformal field theory (AdS$_{d+2}$/CFT$_{d+1}$) correspondence \cite{AdSCFT1,AdSCFT2,AdSCFT3,AdSCFT4}.

After the discovery of the Ryu--Takayanagi formula for the holographic entanglement entropy in the AdS$_{d+2}$/CFT$_{d+1}$ correspondence \cite{RT1,RT2,HRT,RT3}, the multiscale entanglement renormalization ansatz (MERA) \cite{Vidal1,Vidal2} was proposed as the holographic tensor network (HTN) of the bulk quantum entanglement behind this formula in $d=1$ at zero temperature \cite{Swingle1,Swingle2}.
Here, the MERA is the real-space renormalization group transformation of the quantum ground state of the boundary CFT$_2$ in qubits by the semi-infinitely alternate combinations of the layer of the disentanglers (for us, bipartite-qubit gates) and the layer of the coarse-grainers (isometries) \cite{Vidal1,Vidal2}.
The MERA is a scale-invariant tensor network.

Based on the initial studies of the HTN \cite{Swingle1,Matsueda,Review}, the present author formulated the classicalization of the HTN \cite{EPL1,RINP,EPL2,JHAP}.
Here, the {\it classicalization} of the HTN refers to adopting the third Pauli matrix of one-qubits in the HTN as the superselection rule operator \cite{JHAP}.
Namely, the quantum mechanical observables acting on the Hilbert space of the HTN are required to commute with the third Pauli matrix and are selected by this commutativity.
After the classicalization of the HTN, the quantum state of the classicalized holographic tensor network (cHTN) has no quantum interference in the eigenbasis of the third Pauli matrix with respect to the selected observables and thus is equivalent to a classical mixed state, that is, a statistical mixture of product eigenstates of the third Pauli matrix, in the ensemble interpretation of quantum mechanics \cite{dEspagnat}.
The HTN is regarded as a quantum pure or purified state of a Euclidean quantum gravity with holographic ultraviolet completion \cite{Harlow1,Harlow2} in the AdS$_3$/CFT$_2$ correspondence.
There, the Euclidean bulk spacetime with or without a black hole is stationary, and the state of the boundary CFT$_2$ is a thermal equilibrium state including the non-equilibrium steady state (i.e., the thermal and momentum equilibrium state) \cite{EPL1}.
Based on the holographic principle, the present author's main argument was the proposal of the classicalized {\it spin} action of the cHTN \cite{EPL2,JHAP}
\begin{equation}
I_{\rm bulk}[|\psi\ra]=-\hbar bH_{\rm bdy}^{\rm bit}[|\psi\ra]\label{eq:action}
\end{equation}
for the bit factor $b=\ln 2$ and the measurement entropy $H_{\rm bdy}^{\rm bit}$ (i.e., the von Neumann entropy of the classical mixed state of the cHTN as the information lost by the classicalization) of the ground state $|\psi\ra$ of the boundary CFT$_2$.

In Ref.\cite{JHAP}, the ground state of the boundary CFT$_2$ was considered in two independent limits of the large central charge and the strong 't Hooft coupling to derive bulk quantum mechanics from Eq.(\ref{eq:action}).
In the present work, we examine its subtleness motivated by the problem of de Sitter (dS) quantum gravity \cite{dS}\footnote{Recently, there have been developments regarding the dS swampland criterion in any low-energy effective field theory of a consistent quantum gravity; for some notable examples and a review, see Refs.\cite{Sw1,Sw2,Sw3,Sw4,Sw5,Sw6} and Ref.\cite{Sw7}, respectively.}.
In the HTN of the ground state of this CFT$_2$, the measurement entropy of the CFT$_2$ ground state $|\psi\ra$ in bits is
\begin{equation}
H_{\rm bdy}^{\rm bit}[|\psi\ra]=A_{\rm TN}-\alpha_{\rm TN}\;,
\end{equation}
where $A_{\rm TN}$ is the discretized area of the HTN, and positive-valued $\alpha_{\rm TN}$ is the deviation of the measurement entropy from its maximum value $A_{\rm TN}$, that is, the value at the exact strong-coupling limit of the boundary CFT$_2$ \cite{EPL1,EPL2}.
Assuming the scale invariance of the cHTN, we simplify this deviation as a scale-invariant one:
\begin{equation}
\alpha_{\rm TN}=\alpha A_{\rm TN}\label{eq:HMI}
\end{equation}
for a positive-valued number $\alpha$.
This deviation originates from the string dynamics in the bulk spacetime at a finite 't Hooft coupling.
Here, we make two remarks:
\begin{enumerate}
\item[(i)] The deviation term $\hbar b\alpha_{\rm TN}$ in Eq.(\ref{eq:action}) with a real-time duration $T_2$ of the quantum purity of the HTN (i.e., the time constant of the real-time transverse relaxation process of the HTN \cite{NMR}) gives rise to a Lorentzian world-volume action of the HTN as a membrane with a negative tension ${\cal T}_\alpha$ (i.e., a negative inertial mass).

\item[(ii)] The number $\alpha$ arises from looseness of the entangler (i.e., the inverse gate of the disentangler) of a bipartite qubit located at each AdS-scale site of the cHTN, and $\alpha$ is independent of the cHTN scale.
Here, each AdS-scale site bundles the central-charge number of sub-AdS-scale sites, which cannot be distinguished by the quantum state of the cHTN.
\end{enumerate}

The goal of this article is to show that this deviation term $\hbar b\alpha_{\rm TN}$ in the classicalized spin action of the cHTN of the flat spacetime converts the flat spacetime with inexact quantum mechanics in it to the dS$_3$ spacetime with exact quantum mechanics in it by the Wick rotation of the real-time duration $T_2$ of the quantum purity of the HTN.
Here, this Wick rotation keeps the negative tension ${\cal T}_\alpha$ invariant.

As a result, we derive an explicit formula for the positive cosmological constant of this dS$_3$ spacetime and attribute the vanishing smallness of the observed positive cosmological constant to the extraordinary exactness of the quantum mechanics in the flat spacetime.

The rest of this article is organized as follows.
In Sec. II, to introduce some preliminaries, we study the bulk quantum mechanics of a single non-relativistic free particle in the cHTN in the case of $0<\alpha \le 1$.
In this initial work on the subject, to avoid some technical difficulties, we restrict the objects of the bulk quantum mechanics to non-relativistic free particles that have rest masses as matter.
In Sec. III, we give our main statement and its grounds.
In Sec. IV, we conclude this article.

\section{Preliminaries: bulk quantum mechanics}

As with the classicalized spin action (\ref{eq:action}) of the cHTN, we regard the imaginary-time action of a particle, whose dimensions are dropped, as information \cite{JHAP}.
Then, since one DoF (i.e., one-bit information at an AdS-scale site) of the cHTN has the action $\hbar b$, the set of DoF of a non-relativistic free particle reads out an event, $\varepsilon$, from the two {\it bivalent} spin eigenstates of the four bipartite-qubit eigenstates of the third Pauli matrix in the classical mixed state (i.e., a statistical mixture of bivalent spin eigenstates) at a sub-AdS-scale site of the cHTN per its imaginary-time action increment by the amount $\hbar b$ \cite{JHAP}.
Here, the number of events $\varepsilon$ is $W\in [1,2)_{\mathbb R}$ (in the case of $0<\alpha \le 1$) such that
\begin{equation}
W^n=\left(\begin{array}{c}n\\ pn\end{array}\right)\in{\mathbb N}
\end{equation}
for the statistical weight $0\le p \le 1$ of one of the two bivalent spin eigenstates in the classical mixed state at each sub-AdS-scale site of the cHTN holds for $n$ event copies in the large-integer limit of $n$ in the ensemble interpretation of quantum mechanics (e.g., in the case of $\alpha=0$, $p=1/2$ and $W=2$ hold \cite{EPL2}).
We denote the imaginary-time action of the particle by
\begin{equation}
S[\gamma_\tau]=\int_0^\tau d\tau^\prime {\cal H}_{\rm kin}[\gamma_{\tau^\prime}]\label{eq:S}
\end{equation}
with the off-shell trajectory $\gamma_\tau$ of the particle parametrized by the imaginary time $\tau$ with fixed edges and the imaginary-time kinetic Hamiltonian ${\cal H}_{\rm kin}[\gamma_\tau]$ \cite{FH}, and we add $S[\gamma_\tau]$ to the classicalized spin action (\ref{eq:action}) of the cHTN.
Note that, in the case of $\alpha=0$, the classical probability \footnote{Here, {\it classical probability} means probability with no interference.} to obtain an off-shell trajectory $\gamma_\tau$ of the particle with fixed edges is the imaginary-time path integral factor in the exact bulk quantum mechanics
\begin{equation}
W^{-S[\gamma_\tau]/\hbar b}=e^{-S[\gamma_\tau]/\hbar}\;,\ \ W=2\;.
\end{equation}
This equation is the main result of Ref.\cite{JHAP}.

Here, we remark on the rest energy $mc^2$ of a non-relativistic free particle.
\begin{enumerate}
\item[(iii)]
The rest energy makes no contribution to Eq.(\ref{eq:S}).
This stems from the fact that the usual path-integral quantization methods are not consistent for a relativistic {\it particle}.
Even in the description of a relativistic particle in Ref.\cite{FH}, the contributions of rest energy in the exponential factors of the phase and the probability of off-shell trajectories are given by
\begin{equation}
\left\{\begin{array}{cc}{\displaystyle{\frac{-imc^2(t_f-t_i)}{\hbar}}}&{\rm in\ real\ time}\;,\\
& \\
{\displaystyle{\frac{-mc^2(\tau_f-\tau_i)}{\hbar}}}&{\rm in\ imaginary\ time}\;,\end{array}\right.
\end{equation}
respectively.
However, in both cases, the time edges and thus the time interval of the off-shell trajectories are fixed, so these contributions give rise to an {\it overall phase factor} in real time and an {\it extra normalization factor} in imaginary time, respectively.
Thus, even at the fundamental level, the contribution from the rest energy is removed in the following results with respect to non-relativistic free particles.
\end{enumerate}

Next, we denote by $p_{\gamma_{0,N}}^{\rm cl}$ and $\widetilde{p}_{\gamma_{0,N}}^{\rm cl}$ the {\it original} (i.e., $\alpha=0$: $W=2$) and {\it modified} (i.e., $0<\alpha \le 1$: $1\le W<2$) classical probabilities to obtain an off-shell trajectory $\gamma_\tau$ with $N+1$ events and fixed edges, respectively.
Here, $p_{\gamma_{0,N}}^{\rm cl}$ refers to the joint probability
\begin{equation}
p_{\gamma_{0,N}}^{\rm cl}=p[((\gamma_0,{\varepsilon}_0),\tau_0);\ldots;((\gamma_N,{\varepsilon}_N),\tau_N)]
\end{equation}
to obtain the $N+1$ pairs of events with their given imaginary-time parameter values $((\gamma_0,{\varepsilon}_0),\tau_0)$, $\ldots$, $((\gamma_N,{\varepsilon}_N),\tau_N)$, where we set $\tau_0:=0$ and $\tau_N:=\tau$.
We denote the vector of these pairs by
\begin{equation}
\gamma_{0,N}=(((\gamma_0,{\varepsilon}_0),\tau_0),\ldots,((\gamma_N,{\varepsilon}_N),\tau_N))\;.
\end{equation}
The modified classical probability to read out an initial event $(\gamma_1,{\varepsilon}_1)$ at $\tau_1=\tau$ counted from an earlier event $(\gamma_0,{\varepsilon}_0)$ at $\tau_0$ (i.e., $\widetilde{p}_{\gamma_{0,0}}^{\rm cl}=1$) is
\begin{equation}
{\widetilde{p}_{\gamma_{0,1}}^{\rm cl}} = 2^{-(1-\alpha)}=W^{-1}\;.
\end{equation}
This deviates from the imaginary-time path integral factor $p_{\gamma_{0,1}}^{\rm cl}=2^{-1}$ (i.e., $e^{-S[\gamma_{\tau_1}]/\hbar}$) in the exact bulk quantum mechanics by a factor of $2^\alpha$.
As the imaginary-time count of events of the particle
\begin{equation}
N_\tau=\frac{S[\gamma_\tau]}{\hbar b}\label{eq:N}
\end{equation}
grows, the deviation factor $2^\alpha$ grows exponentially as
\begin{equation}
\frac{\widetilde{p}_{\gamma_{0,N_\tau}}^{\rm cl}}{p_{\gamma_{0,N_\tau}}^{\rm cl}}=2^{N_\tau \alpha}\;,\label{eq:key}
\end{equation}
where we set $\widetilde{p}_{\gamma_{0,0}}^{\rm cl}=p_{\gamma_{0,0}}^{\rm cl}=1$.
Now, this relation can be reinterpreted as the exponential contraction of off-shell trajectories of the particle in {\it exact} bulk quantum mechanics.
This is because we postulate the equality between the integrands of the expectation values of the sub-AdS-scale sites of events of the modified vector $\widetilde{\gamma}_{0,N_\tau}$ and the original vector $\gamma_{0,N_\tau}$ for a given set of $N_\tau+1$ imaginary-time parameter values as
\begin{equation}
\widetilde{p}_{\gamma_{0,N_\tau}}^{\rm cl}\widetilde{\gamma}_{0,N_\tau}=p_{\gamma_{0,N_\tau}}^{\rm cl}\gamma_{0,N_\tau}\;,\label{eq:phys}
\end{equation}
and $N_\tau$ in the on-shell trajectory (i.e., the most probable off-shell trajectory), in particular, is proportional to the parameter $\tau$.
Namely, we arrive at the {\it identification}
\begin{equation}
\widetilde{\gamma}_{0,N_\tau}=2^{-N_\tau \alpha}\gamma_{0,N_\tau}\;,\label{eq:result}
\end{equation}
when the bulk quantum mechanics is exact.

\section{Main statement}

Our main statement is that {\it the flat spacetime with inexact quantum mechanics in it is dual to the dS$_3$ spacetime with exact quantum mechanics in it}.
In the exact large-central-charge limit, the flat-space timeslice of the Minkowski spacetime
\begin{eqnarray}
ds^2&=&dy^2+e^{-2y/R_{\rm AdS}}(-c^2dt^2+dx^2)\\
&=&-c^2dt^2+dx^2+dy^2
\end{eqnarray}
for the AdS$_3$ curvature radius $R_{\rm AdS}$ is identified with the flat-space timeslice of a half of the dS$_3$ spacetime
\begin{eqnarray}
ds^2&=&-c^2dt^2+e^{2ct/R_h}(dx^2+dy^2)\\
&=&-c^2dt^2+a_{\rm dS}(t)^2(dr^2+r^2d\theta^2)
\end{eqnarray}
for the radius $R_h$ of the cosmological event horizon satisfying
\begin{equation}
R_h=a_{\rm dS}(t)\int_{t^\prime=t}^\infty dr(t^\prime)\;,\ \ ds=0\;,\ \ d\theta=0
\end{equation}
in the dS$_3$ spacetime \cite{Rindler}.
Here, the duality transformation is given by the Wick rotation of the finite (contracted) real-time duration $T_2$ (i.e., the central charge of the dual CFT$_2$ in the Planck time \cite{EPL2}\footnote{See Ref.\cite{BHe} (respectively, Ref.\cite{dSCFT}) for the AdS$_3$ formula (respectively, the dS$_3$ formula) for the holographic central charge.}) of the quantum purity of the HTN in the world volume with a negative tension ${\cal T}_\alpha$, which is defined in the Lorentzian world-volume action, from the flat phase (a limit of the AdS phase) to the dual dS phase:
\begin{equation}
\frac{\Lambda}{\epsilon^2}=\frac{1}{\left(cT_{2,{\rm dS}}\right)^2}=-\frac{1}{\left(cT_{2,{\rm flat}}\right)^2}\;,\ \ 0<\alpha \le 1\label{eq:Wick}
\end{equation}
for the original cosmological constant $\Lambda$ and a positive infinitesimal $\epsilon$.
Simultaneously, this Wick rotation replaces $\alpha$ with $-i\alpha$ for the invariance of the negative tension
\begin{equation}
{\cal T}_\alpha=-\frac{\hbar b \alpha}{{\cal A}T_{2,{\rm flat}}}<0
\end{equation}
for the finite (contracted) AdS-scale site area ${\cal A}(=|{\cal A}|)$ of the HTN with the finite (contracted) AdS scale $\epsilon R_{\rm AdS}$ given by $R_h$.
Note that the world volume of the HTN with a positive tension
\begin{equation}
{\cal T}_{-1}=\frac{\hbar b}{{\cal A}T_{2,{\rm flat}}}>0
\end{equation}
is in the flat phase (a limit of the AdS phase) (i.e., $T_2=T_{2,{\rm flat}}$).
Here, the flat spacetime has a dual time-contracted (s.t., $t\to \epsilon t$), originally relativistic CFT$_2$ at null infinity (i.e., the conformal boundary) \cite{Flat}.
The null infinity of the flat spacetime is a cylinder, and its CFT$_2$ (a linear combination of two copies of the Virasoro algebra) has two redefined central charges
\begin{equation}
C_1=C-\bar{C}\;,\ \ C_2=\epsilon (C+\bar{C})
\end{equation}
for the central charge $C$ of the original boundary CFT$_2$ and the positive infinitesimal $\epsilon$ \cite{Flat}.
In an originally relativistic CFT$_2$, $C_1=0$ holds.
The finite non-zero redefined central charge $C_2(=C_1+C_2)$ divides the infinitely large total central charge $C+\bar{C}$ of the original boundary CFT$_2$ (two copies of the Virasoro algebra \cite{BHe}) while keeping the geometry of the flat spacetime fixed and contracts the infinitely large AdS-scale site area of the HTN (s.t., $R_h=\epsilon R_{\rm AdS}$), or equivalently, the infinitely long real-time duration of the quantum purity of the HTN (s.t., $T_{2,{\rm flat}}=\epsilon T_{2,{\rm AdS}}$), according to the Brown--Henneaux formula \cite{BHe}
\begin{equation}
C=\bar{C}=\frac{3R_{\rm AdS}}{2G_N}
\end{equation}
for the three-dimensional Newton's gravitational constant $G_N$, thus redefining the discretized area $A_{\rm TN}$ of the HTN (see remark (ii)).

The grounds for this main statement are as follows.
In imaginary time $\tau$, we assume off-shell trajectories of $\nu$ non-relativistic free particles $\gamma_\tau^1,\ldots,\gamma_\tau^\nu$ in the cHTN and consider the original event vector of the center of mass (CM) of the particles
\begin{equation}
\gamma_{0,N_\tau}^{(\rm CM)}=\frac{1}{M}\sum_{i=1}^\nu m_i\gamma_{0,N_\tau}^i\label{eq:CM0}
\end{equation}
for the mass $m_i$ of individual particle $i$ ($i=1,\ldots,\nu$) and the total mass of the particles $M=\sum_{i=1}^\nu m_i$.
In this equation, $N_\tau$ is defined for the imaginary-time action of the CM of the particles $S^{(\rm CM)}[\gamma_\tau^{(\rm CM)}]$ with its off-shell trajectory
\begin{equation}
\gamma_\tau^{(\rm CM)}=\frac{1}{M}\sum_{i=1}^\nu m_i\gamma_\tau^i\;,
\end{equation}
and $\gamma_{0,N_\tau}^i$ ($i=1,\ldots,\nu$) on the right-hand side is not always the original event vector of the $i$-th particle.
Then, the modified event vector of the CM of the particles is given by
\begin{eqnarray}
\widetilde{\gamma}_{0,N_\tau}^{(\rm CM)}&=&\frac{1}{M}\sum_{i=1}^\nu m_i\widetilde{\gamma}_{0,N_\tau}^i\label{eq:gammaCM}\\
&=&\frac{p_{\gamma_{0,N_\tau}^{(\rm CM)}}^{\rm cl}}{\widetilde{p}_{\gamma_{0,N_\tau}^{(\rm CM)}}^{\rm cl}}\gamma_{0,N_\tau}^{(\rm CM)} \\
&=&\frac{1}{M}\sum_{i=1}^\nu m_i\left(\frac{p_{\gamma_{0,N_\tau}^{(\rm CM)}}^{\rm cl}}{\widetilde{p}_{\gamma_{0,N_\tau}^{(\rm CM)}}^{\rm cl}}\gamma_{0,N_\tau}^i\right)\label{eq:CM1}
\end{eqnarray}
via relation (\ref{eq:phys}) by using the modified classical probability $\widetilde{p}_{\gamma_{0,N_\tau}^{(\rm CM)}}^{\rm cl}$ such that
\begin{equation}
\frac{\widetilde{p}_{\gamma_{0,N_\tau}^{(\rm CM)}}^{\rm cl}}{p_{\gamma_{0,N_\tau}^{(\rm CM)}}^{\rm cl}}=2^{N_\tau \alpha}=\exp\left(\frac{S^{(\rm CM)}[\gamma_\tau^{(\rm CM)}]}{\hbar}\alpha\right)
\end{equation}
holds (see remark (iv)).
Namely,
\begin{equation}
\widetilde{\gamma}_{0,N_\tau}^i=\exp\left(-\frac{S^{(\rm CM)}[\gamma_\tau^{(\rm CM)}]}{\hbar}\alpha\right)\gamma_{0,N_\tau}^i
\end{equation}
holds for $i=1,\ldots,\nu$, where $N_\tau$ is defined for the imaginary-time action of the CM of the particles.
From this, we induce the modified off-shell trajectory of the $i$-th particle at $\tau$
\begin{equation}
\widetilde{\gamma}_{\tau^\prime}^i=\exp\left(-\frac{S^{(\rm CM)}[\gamma_\tau^{(\rm CM)}]}{\hbar}\alpha\right)\gamma_{\tau^\prime}^i\;,\ \ 0\le \tau^\prime\le \tau
\end{equation}
for $i=1,\ldots,\nu$.
Then, the original event vector of the $i$-th particle $\gamma_{0,N_\tau}^i$ ($i=1,\ldots,\nu$) is modified to
\begin{eqnarray}
\widetilde{\gamma}_{0,N_\tau}^i=\frac{p_{\gamma_{0,N_\tau}^i}^{\rm cl}}{\widetilde{p}_{\gamma_{0,N_\tau}^i}^{\rm cl}}\gamma_{0,N_\tau}^i
\end{eqnarray}
via relation (\ref{eq:phys}) by using the modified classical probability $\widetilde{p}_{\gamma_{0,N_\tau}^i}^{\rm cl}$ such that
\begin{equation}
\frac{\widetilde{p}_{\gamma_{0,N_\tau}^i}^{\rm cl}}{p_{\gamma_{0,N_\tau}^i}^{\rm cl}}=\exp\left(\frac{S^{(\rm CM)}[\gamma_\tau^{(\rm CM)}]}{\hbar}\alpha\right)
\end{equation}
holds, where $N_\tau$ is defined for the imaginary-time action of the $i$-th particle.
Now, the modification of the original event vector of the CM of the particles (\ref{eq:gammaCM}) in the cHTN is allowed to be reinterpreted as the exponential contraction of the modified scale factor in the Euclidean spacetime
\begin{equation}
\widetilde{a}_{N_\tau}=2^{-N_\tau \alpha}a\;,\label{eq:scale}
\end{equation}
where the bulk quantum mechanics is exact.
Here, $a$ is the original scale factor of unity, and $N_\tau$ is now defined for the on-shell imaginary-time action (i.e., the imaginary-time action evaluated on the classical trajectory) of the CM of the particles in the cHTN.
In real time $t$, after the Wick rotation (\ref{eq:Wick}) (i.e., the duality transformation), the modified scale factor $\widetilde{a}_{{\rm dS},N_t}$ in the dS$_3$ spacetime exponentially expands as
\begin{equation}
\widetilde{a}_{{\rm dS},N_t}=2^{N_t\alpha}a\label{eq:scalefin}
\end{equation}
for the real-time count
\begin{equation}
N_t=iN_\tau\label{eq:Nt}
\end{equation}
of events in the cHTN.
In Eq.(\ref{eq:Nt}), the sign of the kinetic Hamiltonian ${\cal H}_{\rm kin}$ reverses under the Wick rotation from real time $t$ to imaginary time $\tau$.
Equation (\ref{eq:scalefin}) shows a negative tension of the space in real time and is the central result of this article.

Finally, we add the following remark referred to above.
\begin{enumerate}
\item[(iv)] In the case of $\nu\ge 2$, we use relation (\ref{eq:phys}) not for the individual particles as
\begin{equation}
\widetilde{p}_{\gamma_{0,N_\tau}^i}^{\rm cl}\widetilde{\gamma}_{0,N_\tau}^i=p_{\gamma_{0,N_\tau}^i}^{\rm cl}\gamma_{0,N_\tau}^i\;,\ \ i=1,2,\ldots,\nu\;,\label{eq:I}
\end{equation}
where $N_\tau$ in Eq.(\ref{eq:key}) is defined for the imaginary-time actions of the individual particles, respectively, but for the CM of the particles as
\begin{equation}
\widetilde{p}_{\gamma_{0,N_\tau}^{(\rm CM)}}^{\rm cl}\widetilde{\gamma}_{0,N_\tau}^{(\rm CM)}=p_{\gamma_{0,N_\tau}^{(\rm CM)}}^{\rm cl}\gamma_{0,N_\tau}^{(\rm CM)}\;,\label{eq:II}
\end{equation}
where $N_\tau$ in Eq.(\ref{eq:key}) is defined for the imaginary-time action of the CM of the particles.
The reason for this is that relation (\ref{eq:phys}) can be applied to only one set of two DoF by using Eq.(\ref{eq:key}).
If we were to adopt the former (\ref{eq:I}), its elements would contradict each other.
In contrast, from the latter (\ref{eq:II}), relation (\ref{eq:phys}) on each particle follows.
The adoption of Eq.(\ref{eq:II}) is not a mathematical consequence but a physical postulate, which is based on the conservation of total momentum, as with relation (\ref{eq:phys}) itself.

\end{enumerate}

\section{Conclusion}

In our scenario, the positive cosmological constant of the dS$_3$ spacetime with exact quantum mechanics in it is in the second order of the degree $\alpha$ of the violation of quantum mechanics in the dual flat spacetime (refer to remark (ii)):
\begin{equation}
\Lambda_{\rm dS}\sim \frac{\alpha^2}{\left(ct_{\rm ML}\right)^2}\label{eq:CC}
\end{equation}
for the Margolus--Levitin time (i.e., the minimum time required to rotate a state vector to its orthogonal state vector quantum mechanically) \cite{ML}
\begin{equation}
t_{\rm ML}=\frac{h}{4E}
\end{equation}
defined for kinetic energy $E$ (see remark (iii)) of the CM of the non-relativistic free particles in the original on-shell trajectory in the flat spacetime, which depends on the choice of inertial reference frame (i.e., inertial observer), due to Eqs.(\ref{eq:S}), (\ref{eq:N}), and (\ref{eq:scalefin}).
From calculations in Ref.\cite{Lloyd}, the Margolus--Levitin time $t_{\rm ML}$ defined for $E$ in the matter-dominated Universe is estimated as
\begin{equation}
t_{\rm ML}\sim 10^{-102} \left(\frac{c}{v}\right)^2\ [{\rm s}]\label{eq:Lloyd}
\end{equation}
for speed $v$ of the CM of the non-relativistic free particles.
If we can extend our scenario to the dS$_4$ spacetime, then Eqs.(\ref{eq:CC}) and (\ref{eq:Lloyd}) suggest that the vanishing smallness of the observed positive cosmological constant \cite{CC}
\begin{equation}
\Lambda_{\rm dS}^{\rm obs}\sim 10^{-122}(ct_P)^{-2}
\end{equation}
for the Planck time $t_P\sim 10^{-43}$ [s] is attributable to the extraordinary exactness of the quantum mechanics in the flat spacetime.
Such a statement is possible because the classicalized spin action (\ref{eq:action}) of the cHTN, that is, $\hbar b$ times the negative amount of information of the cHTN in bits, does not require the minimum action principle but requires the principal argument that the most probable configuration of the cHTN (i.e., the highest measurement entropy of the cHTN) is likely realizable.

\end{document}